\newlength{\absize}
\documentclass[12pt]{article}
\usepackage{graphicx}
\usepackage{latexsym}
\usepackage{amssymb}
\usepackage{bbm}
\usepackage{setspace}
\usepackage{braket}
\renewcommand{\arraystretch}{2}
\newcommand{\tr}{\mathop{\rm Tr}\nolimits}
\newcommand{\figsize}{\small}

\renewcommand{\bar}{\overline}

\newcommand{\spur}[1]{\!\not\! #1 \,}

\newcommand{\cA}{\mathcal{A}}
\newcommand{\cV}{\mathcal{V}}

\newcommand{\cL}{\mathcal{L}}

\newcommand{\cB}{\mathcal{B}}
\newcommand{\cC}{\mathcal{C}}

\newcommand{\bA}{\mathbbm{A}}
\newcommand{\bV}{\mathbbm{V}}

\newcommand{\pd}{\partial}

\renewcommand{\slash}[1]{#1\!\!\!/}

\newcommand{\be}{\begin{equation}}
\newcommand{\ee}{\end{equation}}
\newcommand{\bea}{\begin{eqnarray}}
\newcommand{\eea}{\end{eqnarray}}

\newcommand{\comment}[1]{}

\begin{document}

\thispagestyle{empty}
\pagestyle{empty}
\newcommand{\starttext}{\newpage\normalsize
 \pagestyle{plain}
 \setlength{\baselineskip}{3ex}\par
 \setcounter{footnote}{0}
 \renewcommand{\thefootnote}{\arabic{footnote}}
 }
\newcommand{\preprint}[1]{\begin{flushright}
 \setlength{\baselineskip}{3ex}#1\end{flushright}}
\renewcommand{\title}[1]{\begin{center}\LARGE
 #1\end{center}\par}
\renewcommand{\author}[1]{\vspace{2ex}{\large\begin{center}
 \setlength{\baselineskip}{3ex}#1\par\end{center}}}
\renewcommand{\thanks}[1]{\footnote{#1}}
\renewcommand{\abstract}[1]{\vspace{2ex}\normalsize\begin{center}
 \centerline{\bf Abstract}\par\vspace{2ex}\parbox{\absize}{#1
 \setlength{\baselineskip}{2.5ex}\par}
 \end{center}}

\title{Generalizations of the Sommerfield and Schwinger models}
\author{
 Howard~Georgi\thanks{\noindent \tt hgeorgi@fas.harvard.edu}
and
Brian~Warner\thanks{\noindent \tt brian\_warner@college.harvard.edu} 
\\ \medskip
Center for the Fundamental Laws of Nature\\
Jefferson Physical Laboratory \\
Harvard University \\
Cambridge, MA 02138
 }
\date{\today}
\abstract{The Sommerfield model with a massive vector field coupled to a
massless fermion in 1+1 dimensions is an exactly solvable analog of a
Bank-Zaks model.  The ``physics'' of the model comprises a massive boson
and an unparticle
sector that survives at low energy as a conformal field theory (Thirring
model).  
We analyze generalizations of the Sommerfield model, and the corresponding
generalizations of the Schwinger model, with more massless
fermions and more
vector fields.
}

\starttext

\section{Generalized Sommerfield model of a Banks-Zaks sector\label{sec-model}}

The study of explicitly solvable field theories in 1+1 dimensions has
been a fruitful source of important ideas about quantum field theories in
general~\cite{Schwinger:1962tp, Wilson:1969zs, Coleman:1976uz}.   
In this note, we hope to contribute to this rich legacy by 
constructing and analyzing generalizations of
the Schwinger and Sommerfield~\cite{Sommerfield:1964}
models
with multiple massless fermions and vector fields.\footnote{Our conventions, as
in~\cite{Georgi:2008pq}, are:  
$
g^{00}=-g^{11}=1,\,
\epsilon^{01}=-\epsilon^{10}=-\epsilon_{01}=\epsilon_{10}=1
$. From the defining properties $\{\gamma^\mu,\gamma^\nu\} = 2g^{\mu\nu}$ 
and $\gamma^5 = -\frac{1}{2}\epsilon_{\mu\nu}\gamma^\mu\gamma^\nu$, it follows that
$\gamma^\mu\gamma^5=-\epsilon^{\mu\nu}\gamma_\nu$ and
$\gamma^\mu\gamma^\nu=g^{\mu\nu}+\epsilon^{\mu\nu}\gamma^5$, and we will
use the representation 
$
\gamma^0=
\pmatrix{
0&1\cr
1&0\cr
},\;
\gamma^1=
\pmatrix{
0&-1\cr
1&0\cr
},\;
\gamma^5=\gamma^0\gamma^1=
\pmatrix{
1&0\cr
0&-1\cr
}\,$. Then the components $\psi_1$ and $\psi_2$ describe right-moving and
left-moving fermions, respectively.
Lightcone coordinates are defined by
$$x^\pm={x^0\pm x^1}
\quad\quad
\partial_\pm=\frac{\partial_0\pm\partial_1}{2}$$
$$x^+\partial_++x^-\partial_-=(x^0+x_1)\frac{\partial_0+\partial_1}{2}+(x^0-x_1)\frac{\partial_0-\partial_1}{2}
=x^0\partial_0+x^1\partial_1$$
$$
A^0=\pd^0\cV/m_0-\pd^1\cA/m
\quad\quad
A^1=\pd^1\cV/m_0-\pd^0\cA/m
$$
$$
A_0=\pd_0\cV/m_0+\pd_1\cA/m
\quad\quad
A_1=\pd_1\cV/m_0+\pd_0\cA/m
\quad\quad
A_\pm=\pd_\pm\cV/m_0\pm\pd_\pm\cA/m
$$
}
In this note, we find the general fermion correlation functions for an
arbitrary number of massless fermions with commuting couplings to an
arbitrary number vector fields in 1+1 dimensions.  While the general
correlators including the vector fields are all calculable using the
results we find in this paper, we focus on the fermion correlators which
are particularly interesting because they determine the long-distance
physics.  A number of interesting special cases have appeared in the
literature, see for example \cite{Lowenstein:1971fc}, \cite{Rothe:1976kn},
and \cite{Belvedere:1978fj}.

In \cite{Georgi:2019vyk}, one of us suggested that it useful to think of
the Schwinger model as a limit of the Sommerfield model as the gauge bare
vector meson
mass goes to zero.  In a separate note~\cite{Georgiandwarner2}, we will use the technology
developed in this paper to construct and analyze a number of generalized
Schwinger models.

We will begin with a quick review of the Sommerfield model to set
notation.\footnote{The analysis is similar to that in \cite{Georgi:2009xq}
and inspired by the treatment of the Schwinger model in \cite{Lowenstein:1971fc}.}
The Sommerfield Lagrangian is
\be
\cL_S =
\bar\psi\,(i\spur\pd - e\slash A)\,\psi
-\frac{1}{4}F^{\mu\nu} F_{\mu\nu}
+\frac{m_0^2}{2}A^\mu\, A_\mu
\label{Sommerfield-model2}
\ee
To solve the model, we decompose $A^\mu$ as
\be
A^\mu = \pd^\mu \cV/m_0 + \epsilon^{\mu\nu}\pd_\nu\cA/m
\label{A-decomposition}
\ee
where
\begin{equation}
m^2=m_0^2+e^2/\pi
\end{equation}
Then we can write
\begin{equation}
\epsilon_{\mu\nu}\partial^\mu A^\nu=
\epsilon_{\mu\nu}\partial^\mu\epsilon^{\nu\beta}\pd_\beta\cA/m
=\partial_\mu\partial^\mu\cA/m
\quad\quad
\partial_\mu A^\mu=\partial_\mu\partial^\mu\cV/m_0
\end{equation}
and the Lagrangian becomes
\be
\displaystyle
\cL_S = i\bar\psi\spur\pd\psi -
e\bar\psi\gamma_\mu\psi\left(\pd^\mu\cV/m_0 +
\epsilon^{\mu\nu}\pd_\nu\cA/m\right)
 + \frac{1}{2m^2}\cA\,\Box^2\cA +
\frac{1}{2}\pd_\mu\cV\pd^\mu\cV -\frac{m_0^2}{2m^2}
\pd_\mu\cA\pd^\mu\cA
\label{Sommerfield-AV}
\ee

If we change the fermionic variable to
\be
\Psi = e^{ie\left(\cV/m_0 + \cA\gamma^5/m\right)}\psi
\label{psi-redef}
\ee
the fermion becomes free:
\be
\cL_S=i\bar\Psi\spur\pd\Psi + \frac{1}{2}\pd_\mu\cV\pd^\mu\cV +
\frac{1}{2m^2}\cA\,\Box^2\cA - \frac{1}{2}\pd_\mu\cA\pd^\mu\cA
\label{Sommerfield-redefined}
\ee

While the final result is expressed in terms of free fields, it is still
worth being careful about the definitions of composite operators, with
several local fields at the same point.  Because of the
super-renormalizable interactions, it is not much of an issue in these
theories.  Almost all the relevant operators can be defined simply as the
leading operators in an operator product expansion~\cite{Wilson:1970pq}.
However, there is one important exception.
In the last terms in (\ref{Sommerfield-redefined}), 
$m_0^2/m^2$ has been replaced
by 1
in order to account for the fact that the path integral measure is not
invariant under the $\cA$ part of (\ref{psi-redef})~\cite{Roskies:1980jh}.\footnote{The same
effect gives mass $e/\sqrt\pi$ to the gauge boson in the Schwinger
model. See also~\cite{Georgi:1971iu}.}  This effect can be seen in many
different ways.

It is worth recalling how this works in more detail.
We can get from the kinetic energy in (\ref{Sommerfield-AV}) 
\be
 \bar\psi\,\gamma_\mu\bigl(i\pd^\mu
 -e\left(\pd^\mu\cV/m_0 +
\epsilon^{\mu\nu}\pd_\nu\cA/m\right)\bigr)\,
\psi
\label{Sommerfield-AV-ke}
\ee
to
\be
 \overline{e^{ie\left(\cV/m_0 \right)}\psi}\,\gamma_\mu\bigl(i\pd^\mu
 -e\,
\epsilon^{\mu\nu}\pd_\nu\cA/m\bigr)\,
e^{ie\left(\cV/m_0 \right)}\psi
\label{Sommerfield-AV-ke-v}
\ee
by an ordinary vector gauge transformation, so (\ref{Sommerfield-AV-ke})
and  (\ref{Sommerfield-AV-ke-v}) are equal.
But $\pd^\mu\cA$ has
axial-vector couplings because 
\begin{equation}
\gamma_\mu
\epsilon^{\mu\nu}\pd_\nu\cA
=\gamma_\mu\gamma^5\pd^\mu\cA
\end{equation}
and an axial gauge transformation induces a change in the Lagrangian.   The
effect from an infintesimal axial transformation is proportional to the 2D
axial anomaly,
\begin{equation}
\pd_\mu j_5^\mu=-\frac{e}{\pi}\epsilon^{\mu\nu}\pd_\mu A_\nu=-\frac{e}{\pi}\Box\cA
\label{anomaly}
\end{equation}
\be
\frac{d}{d\alpha}
 \overline{e^{ie\left(\cV/m_0+\alpha\cA\gamma^5/m \right)}\psi}\,\gamma_\mu\bigl(i\pd^\mu
 -e\,
\epsilon^{\mu\nu}\pd_\nu(1-\alpha)\cA/m\bigr)\,
e^{ie\left(\cV/m_0+\alpha\cA\gamma^5/m \right)}\psi
=-\frac{e^2}{m^2\pi}\cA(1-\alpha)\Box\cA
\label{Sommerfield-AV-ke-vda}
\ee
Integrating (\ref{Sommerfield-AV-ke-vda}) from $\alpha=0$ to $1$ gives
\begin{equation}
 \overline{e^{ie\cV/m_0}\psi}\,\gamma_\mu\bigl(i\pd^\mu
 -e\,
\epsilon^{\mu\nu}\pd_\nu\cA/m\bigr)\,
e^{ie\cV/m_0}\psi
=i\bar\Psi\spur\pd\Psi
+\frac{e^2}{2m^2\pi}\cA\Box\cA
=i\bar\Psi\spur\pd\Psi
-\frac{e^2}{2m^2\pi}\pd_\mu\cA\pd^\mu\cA
\label{anomaly2}
\end{equation}
where $\Psi$
is given by (\ref{psi-redef}).  The second term in (\ref{anomaly2})
produces the change in the coefficient of the $\cA$ kinetic energy to its
canonical value.

Focusing on $\cA$ in (\ref{Sommerfield-redefined}), we can replace it with somewhat more normal looking fields
as follows.
\be
\frac{1}{2m^2}\cA\,\Box^2\cA - \frac{1}{2}\pd_\mu\cA\pd^\mu\cA
\to
-\frac{m^2}{2}\cB^2+\cB\Box\cA - \frac{1}{2}\pd_\mu\cA\pd^\mu\cA
\ee
\be
=-\frac{m^2}{2}\cB^2+\frac{1}{2}\pd_\mu\cB\pd^\mu\cB
- \frac{1}{2}\pd_\mu\cC\pd^\mu\cC
\ee
where $\cC=\cA+\cB$, so $\cB$ is a massive field
and $\cC$ is a massless ghost and
the Lagrangian becomes
\be
\cL_S=i\bar\Psi\spur\pd\Psi + \frac{1}{2}\pd_\mu\cV\pd^\mu\cV 
-\frac{m^2}{2}\cB^2+\frac{1}{2}\pd_\mu\cB\pd^\mu\cB
- \frac{1}{2}\pd_\mu\cC\pd^\mu\cC\\
\label{Sommerfield-redefined2}
\ee
and the original fermion and vector fields can be written in terms of free fields
\be
\psi_S = e^{-ie\left(\cV/m_0 + (\cC-\cB)\gamma^5/m\right)}\Psi
\label{psi-redef-vcb}
\ee
\be
A_S^\mu = \pd^\mu \cV/m_0 + \epsilon^{\mu\nu}\pd_\nu(\cC-\cB)/m
\label{avbc}
\ee

In the following sections, we generalize this analysis to include more
fermions and more vector bosons.
We can then use the generalizations of (\ref{psi-redef-vcb}) and (\ref{avbc}) to
write down the Green's functions as in~\cite{Georgi:2009xq}.  This is done in section~\ref{sec-g1}.

\section{$n_F$ fermions\label{sec-fermions}}

We now need an index, $\alpha=1$ to $n_F$ for the fermions, and we can
write the Lagrangian as
\be
\cL_f =
\left( \sum_{\alpha=1}^{n_F}\bar\psi_\alpha\,(i\spur\pd - [e]_\alpha\slash A)\,\psi_\alpha\right)
-\frac{1}{4}F^{\mu\nu} F_{\mu\nu}
+\frac{m_0^2}{2}A^\mu\, A_\mu
\label{f-Sommerfield-model2}
\ee
We have written the coupling in a peculiar way because in further
generalizations, it will be convenient to use a matrix notation and
think of the couplings as forming
a vector, $e$, in the fermion space, so $[e]_\alpha$ is just the
$\alpha$ component of $e$.
Again, we decompose $A^\mu$ as
\be
A^\mu = \pd^\mu \cV/m_0 + \epsilon^{\mu\nu}\pd_\nu\cA/m
\label{f-A-decomposition}
\ee
where
\begin{equation}
m^2=m_0^2+e\,e^T/\pi
\label{f-m-m0}
\end{equation}
because each of the massless fermions contributes to dynamical mass of the
vector field.
Thus (\ref{f-Sommerfield-model2}) becomes
\be
\begin{array}{c}
\displaystyle
\cL_f =\left(\sum_{\alpha=1}^{n_F}\Bigl( i\bar\psi_\alpha\spur\pd\psi_\alpha -
e_\alpha\bar\psi_\alpha\gamma_\mu\psi_\alpha\left(\pd^\mu\cV/m_0 +
\epsilon^{\mu\nu}\pd_\nu\cA/m\right)\Bigr)\right)
\\ \displaystyle
 + \frac{1}{2m^2}\cA\,\Box^2\cA +
\frac{1}{2}\pd_\mu\cV\pd^\mu\cV -\frac{m_0^2}{2m^2}
\pd_\mu\cA\pd^\mu\cA
\end{array}
\label{f-Sommerfield-AV}
\ee
If we change the fermionic variables to
\be
\Psi_\alpha = e^{i[e]_\alpha\left(\cV/m_0 + \cA\gamma^5/m\right)}\psi
\label{f-psi-redef}
\ee
the fermions becomes free:
\be
\cL_f=\left(\sum_{\alpha=1}^{n_F}i\bar\Psi_\alpha\spur\pd\Psi_\alpha\right) + \frac{1}{2}\pd_\mu\cV\pd^\mu\cV +
\frac{1}{2m^2}\cA\,\Box^2\cA - \frac{1}{2}\pd_\mu\cA\pd^\mu\cA
\label{f-Sommerfield-redefined}
\ee
As usual, in the last terms in (\ref{f-Sommerfield-redefined}) 
$m_0^2/m^2$ has been replaced
by 1
in order to account for the fact that the path integral measure is not
invariant under the $\cA$ part of (\ref{f-psi-redef})~\cite{Roskies:1980jh}.

Again we can replace
$\cA$ in (\ref{f-Sommerfield-redefined}) with somewhat more normal looking fields
as follows.
\be
\frac{1}{2m^2}\cA\,\Box^2\cA - \frac{1}{2}\pd_\mu\cA\pd^\mu\cA
\to
-\frac{m^2}{2}\cB^2+\cB\Box\cA - \frac{1}{2}\pd_\mu\cA\pd^\mu\cA
\ee
\be
=-\frac{m^2}{2}\cB^2+\frac{1}{2}\pd_\mu\cB\pd^\mu\cB
- \frac{1}{2}\pd_\mu\cC\pd^\mu\cC
\ee
where $\cC=\cA+\cB$, so $\cB$ is a massive field
and $\cC$ is a massless ghost.
Then
\be
\cL_f=\left(\sum_{\alpha=1}^{n_F}i\bar\Psi_\alpha\spur\pd\Psi_\alpha\right)
 + \frac{1}{2}\pd_\mu\cV\pd^\mu\cV 
-\frac{m^2}{2}\cB^2+\frac{1}{2}\pd_\mu\cB\pd^\mu\cB
- \frac{1}{2}\pd_\mu\cC\pd^\mu\cC\\
\label{f-Sommerfield-redefined2}
\ee
and the original fermion and vector fields can be written in terms of free fields
\be
\psi_\alpha = e^{-i[e]_\alpha\left(\cV/m_0 + (\cC-\cB)\gamma^5/m\right)}\Psi_\alpha
\label{f-psi-redef-vcb}
\ee
and as usual
\be
A^\mu = \pd^\mu \cV/m_0 + \epsilon^{\mu\nu}\pd_\nu(\cC-\cB)/m
\label{f-avbc}
\ee
As usual, we can use  (\ref{f-psi-redef-vcb}) and (\ref{f-avbc})
straightforwardly to write down the Green's functions
as in~\cite{Georgi:2009xq}.
The long distance physics of (\ref{f-Sommerfield-redefined2}) is a conformal
field theory and the fermion fields are conformal operators.

\section{$n_A$ Vectors\label{sec-n-vectors}}

The analysis also generalizes in a straightforward way to more vector fields.
Our generalized Lagrangian is
\be
\cL_A =
\bar\psi\,(i\spur\pd - e^T\slash A)\,\psi
-\frac{1}{4}F^{T\mu\nu}\cdot F_{\mu\nu}
+\frac{1}{2}A^{\mu T}\,M_0^2\, A_\mu
\label{n-Sommerfield-model2m}
\ee
where our vector fields $A$ and couplings $e$ are $n_A$ dimensional column
vectors, and $M_0^2$ is a positive $n_A\times n_A$ matrix.\footnote{Later, we
will relax this condition and consider what happens when $m_0^2$ has a zero
eigenvalue, but we can approach this interesting case as a limit.}
Where it does not cause confusion, we may drop the transposes and not
distinguish between row vectors and column vectors, so we can write
\be
\cL_A =
\bar\psi\,(i\spur\pd - e\cdot\slash A)\,\psi
-\frac{1}{4}F^{\mu\nu}\cdot F_{\mu\nu}
+\frac{1}{2}A^\mu\,M_0^2\, A_\mu
\label{n-Sommerfield-model2}
\ee
As in the Sommerfield model,
to solve the model, it is convenient to decompose $A^\mu$ into scalar and
pseudo-scalar fields.  Because the mass structure is now more complicated,
we will do this in two steps.  Begin with
\be
A^\mu = \pd^\mu \bV + \epsilon^{\mu\nu}\pd_\nu\bA
\label{n-A-decomposition0}
\ee
The Lagrangian becomes
\be
\cL = i\bar\psi\spur\pd\psi -
\bar\psi\gamma_\mu\psi\,e\cdot\left(\pd^\mu\bV +
\epsilon^{\mu\nu}\pd_\nu\bA\right) + \frac{1}{2}\bA\,\cdot\Box^2\bA +
\frac{1}{2}\left(\pd_\mu\bV\,M_0^2\,\pd^\mu\bV -
\pd_\mu\bA\,M_0^2\,\pd^\mu\bA\right)
\label{n-Sommerfield-AV}
\ee
If we change the fermionic variable to
\be
\Psi = e^{ie\cdot(\bV + \bA\gamma^5)}\psi
\label{n-psi-redef0}
\ee
the fermion becomes free:
\be
\cL = i\bar\Psi\spur\pd\Psi + \frac{1}{2}\pd_\mu\bV\,M_0^2\,\pd^\mu\bV +
\frac{1}{2}\bA\cdot\Box^2\bA - \frac{1}{2}\pd_\mu\bA\,M^2\,\pd^\mu\bA 
\label{n-Sommerfield-redefined}
\ee
In the last term of (\ref{n-Sommerfield-redefined}) $M_0^2$ has been replaced
by
\be
M^2 = M_0^2 + \frac{1}{\pi}\,e\,e^T
\label{m2extra}
\ee
This is the same physics that we saw in (\ref{Sommerfield-redefined})
in order to account for the fact that the path integral measure is not
invariant under the $\cA$ part of
(\ref{n-psi-redef})~\cite{Roskies:1980jh}.

So far, we are considering the case in which
both $M_0^2$ and $M^2$ are
positve matrices, 
so we can define the positive square
roots $M_0$ and $M$  
and use these to simplify things by taking
\begin{equation}
\cV=M_0\bV
\quad\mbox{and}\quad
\cA=M\bA
\end{equation}
and write the original $\psi$ and $A^\mu$ fields in the Lagrangian in terms
of free fields, as in the Sommerfield model.
\be
\psi = e^{-ie\cdot(M_0^{-1}\cV + M^{-1}\cA\gamma^5)}\Psi
\label{n-psi-redef}
\ee
\be
A^\mu = M_0^{-1}\pd^\mu \cV + M^{-1}\epsilon^{\mu\nu}\pd_\nu\cA
\label{n-A-decomposition}
\ee

Now the $\cV$ kinetic energy is conventional and as in 
section~\ref{sec-n-vectors}, we can
replace $\cA$ with more normal-looking fields ---
\be
\begin{array}{c}
\displaystyle
\frac{1}{2}\cA\,M^{-2}\Box^2\cA - \frac{1}{2}\pd_\mu\cA\cdot\pd^\mu\cA
\to
-\frac{1}{2}\cB\,M^2\,\cB+\cB\cdot\Box\cA - \frac{1}{2}\pd_\mu\cA\cdot\pd^\mu\cA\\
\displaystyle
=-\frac{1}{2}\cB\,M^2\,\cB+\frac{1}{2}\pd_\mu\cB\cdot\pd^\mu\cB
- \frac{1}{2}\pd_\mu\cC\cdot\pd^\mu\cC
\end{array}
\ee
where $\cC=\cA+\cB$. Then $\cB$ are massive fields and $\cC$ are massless
ghosts.   
And
\be
\psi = e^{-ie\cdot(M_0^{-1}\cV + M^{-1}(\cC-\cB)\gamma^5)}\Psi
\label{n-psi-redef-bc}
\ee
\be
A^\mu = M_0^{-1}\pd^\mu \cV + M^{-1}\epsilon^{\mu\nu}\pd_\nu(\cC-\cB)
\label{n-A-decomposition-bc}
\ee

We are free to make an orthogonal
transformation on the vector fields which induces an orthogonal
transformation on the $e$ matrix,
\begin{equation}
e\to O\,e
\label{oe}
\end{equation}
where $O$ is a real orthogonal $n_A\times n_A$ matrix.
We can use this freedom to 
take the physical vector boson mass matrix $M$ to be diagonal
and write
\begin{equation}
[M]_{jk}=m_j\,\delta_{jk}
\quad\quad
[M_0^2]_{jk}=m_j^2\,\delta_{jk}-\frac{[e]_j\,[e]_k}{\pi}
\label{n-pi}
\end{equation}
Then the couplings of the $\cB$ and $\cC$ are simple, and all the
complication is in the couplings of the $\cV$.
\be
\psi = e^{-ie\cdot(M_0^{-1}\cV + M^{-1}(\cC-\cB)\gamma^5)}\Psi
= e^{-i(e\cdot M_0^{-1}\cV + \sum_j(e_j/m_j)([\cC]_j-[\cB]_j)\gamma^5)}\Psi
\label{n-psi-redef-bc-d}
\ee
\be
[A^\mu]_j = [M_0^{-1}\pd^\mu \cV]_j +\frac{1}{m_j}\epsilon^{\mu\nu}\pd_\nu([\cC]_j-[\cB]_j)
\label{n-A-decomposition-bc-d}
\ee

Notice that (\ref{n-pi}) and the positivity of
$M_0^2$ implies that if $m_j=0$, then $[e]_j$ must be zero as well, so $A_j$
decouples and is not interesting.  So we are not interested in the
situtation in which any of the $m_j$ go to zero and will assume that they
are all positive. 

We can use (\ref{n-pi}) to simplify the calculation of the
$M_0$ dependence of the correlators.  Note that
(\ref{n-pi}) can be written as
\begin{equation}
M_0^2=M^2-\frac{1}{\pi}\,e\,e^T
=M\left(I-\frac{1}{\pi}M^{-1}e\,e^TM^{-1}\right)M
\label{n-pi-braket}
\end{equation}
This is positive so long as the eigenvalues of the matrix
\begin{equation}
\tilde B_1\equiv \frac{1}{\pi}M^{-1}e\,e^TM^{-1}
\label{betamatrix}
\end{equation}
are less than 1. The matrix $\tilde B_1$ is rank-1 and all but one of the eigenvalues
vanish, so the non-zero eigenvalue is just the trace
\begin{equation}
\beta=\tr\tilde B_1
=\frac{1}{\pi}e^TM^{-2}e
=\frac{1}{\pi}\sum_j\,\frac{[e]_j^2}{m_j^2}
\label{beta}
\end{equation}
Thus we lose positivity of $M_0^2$ if and only if $\beta>1$ so we must have
\begin{equation}
\sum_j\frac{e_j^2}{m_j^2}\leq\pi
\label{n-constraint}
\end{equation}
The equality corresponds to the interesting case in which one of the eigenvalues of
$M_0$ goes to zero.
This is the ``Schwinger point''~\cite{Georgi:2019vyk} at which one linear combination of the
vectors has zero mass so there is a gauge invariance.
We will discuss that further in \cite{Georgiandwarner2}.

But going on, we can now formally write down $M_0^{-2}$, in the form
\begin{equation}
M_0^{-2}
=M^{-1}\left( I+\frac{\gamma}{\pi}M^{-1}e\,e^TM^{-1} \right)M^{-1}
\label{gamma}
\end{equation}
Then $\gamma$ must satisfy
\begin{equation}
M_0^{-2}M_0^2=I
=M^{-1}\left(I+\frac{-1+\gamma-\beta\gamma}{\pi}M^{-1}e\,e^TM^{-1}\right)M
\label{gamma-condition}
\end{equation}
so
\begin{equation}
\gamma=\frac{1}{1-\beta}
\label{gamma2}
\end{equation}
and
\begin{equation}
M_0^{-2}
=M^{-1}\left(I+\frac{1}{\pi(1-\beta)}M^{-1}e\,e^TM^{-1}\right)M^{-1}
\label{m0-2gamma}
\end{equation}

In the fermion correlators, from (\ref{n-psi-redef-bc-d}), because the
$\cV$ are all massless, the mass matrix $M_0$ only appears in
the combination $e^TM_0^{-2}\,e$ which from (\ref{m0-2gamma}) is 
\begin{equation}
e^TM_0^{-2}\,e
=e^TM^{-1}\left(I+\frac{1}{\pi(1-\beta)}M^{-1}e\,e^TM^{-1}\right)M^{-1}e
=\frac{\pi\beta}{1-\beta}
\label{m0inC}
\end{equation}

We focus in this note on the fermion correlators, but it is amusing to note
that
to calculate the correlations functions involving fermions and a single
$A_j$, the only additional thing we need to calculate that
depends on $M_0$ is
$M_0^{-2}e$.  A factor of $M_0^{-1}e$ comes from the $\cV$
dependence of the fermion
field --- and this is complicated.  But this gets multiplied by an
additional factor of $M_0^{-1}$ from the $A_j$ field, and the combination
is simple:
\begin{equation}
M_0^{-2}e
=M^{-1}\left(I+\frac{1}{\pi(1-\beta)}M^{-1}e\,e^TM^{-1}\right)M^{-1}e
=\frac{1}{1-\beta}M^{-2}e
\label{m0inAj}
\end{equation}

The ghost contribution is just proportional, with the coefficient
\begin{equation}
e^TM^{-2}e=\pi\beta
\end{equation}
This is the combination that appears in the unparticle part of the $C$ functions.  At the
Schwinger point, it just goes to $\pi$.

\section{$n_A$ vectors and $n_F$ fermions\label{sec-na-vectors-nf-fermions}}

This is the general case, so in this
section, we have an index for the fermions, $\psi_\alpha$ for $\alpha=1$ to
$n_F$,
an index for the vector bosons, $A^\mu_j$ for $j=1$ to $n_A$, and
$n_F$ sets
of couplings, one to each of the fermions. So in a generalization of the notation of the previous
section, we can write the couplings as a $n_F\times n_A$ matrix, $e$, in which
matrix in which
\begin{equation}
[e]_{j\alpha}\mbox{~is the coupling of the $j$th vector to the $\alpha$th fermion.}
\label{efa}
\end{equation}
Where the $\ket{j}$ are an orthonormal basis in the index space of the
vector fields.  We will often think of $e$ as a  and use a matrix notation.  The Lagrangian is
\be
\cL_{Af} =
\left(\sum_{\alpha=1}^{n_F}\bar\psi_\alpha\,(i\spur\pd - [e^T\not\!\!A]_\alpha)\,\psi_\alpha\right)
-\frac{1}{4}F^{\mu\nu}\cdot F_{\mu\nu}
+\frac{1}{2}A^\mu\,M_0^2\, A_\mu
\label{vfn-Sommerfield-model2}
\ee
Note that we have assumed that the gauge couplings are diagonal in the
fermion space.  This is important to ensure that the model is exactly
solvable.  We could have non-diagonal couplings as long as the couplings to
different gauge bosons commute with one another.  But then we can
simultaneously diagonalize them with a unitary transformation on the
fermion fields, so we may as well assume that the gauge couplings are
diagonal from the beginning.  

As one might expect, the analysis of this model uses a combination of the
tools discussed in the two previous sections.
Each of the massless fermions generates a contribution to the vector boson
mass matrix like that in (\ref{m2extra}), 
\begin{equation}
M^2=M_0^2+\frac{1}{\pi}e\,e^T
\end{equation}
As in section~\ref{sec-n-vectors}, we will use our freedom to redefine the
vector fields to diagonalize the physical vector boson mass, so as in
(\ref{n-pi}), we can write
and write
\begin{equation}
[M]_{jk}=m_j\,\delta_{jk}
\quad\quad
[M_0^2]_{jk}=m_j^2\,\delta_{jk}-\sum_\alpha\frac{[e]_{j\alpha}\,[e]_{k\alpha}}{\pi}
\label{n-pi-alpha}
\end{equation}
Then as in section~\ref{sec-n-vectors}, the couplings of the $\cB$ and $\cC$ are simple, and all the
complication is in the couplings of the $\cV$.
\be
\psi_\alpha = e^{-i\bigl([e^T\cdot M_0^{-1}\cV]_\alpha + [e^TM^{-1}(\cC-\cB)]_\alpha\gamma^5\bigr)}\Psi_\alpha
= e^{-i\bigl([e^T\cdot M_0^{-1}\cV]_\alpha + \sum_j([e]_{j\alpha}/m_j)([\cC]_j-[\cB]_j)\gamma^5\bigr)}\Psi_\alpha
\label{n-psi-redef-bc-d-alpha}
\ee
\be
[A^\mu]_j = [M_0^{-1}\pd^\mu \cV]_j +\frac{1}{m_j}\epsilon^{\mu\nu}\pd_\nu([\cC]_j-[\cB]_j)
\label{n-A-decomposition-bc-d-alpha}
\ee
The relations
(\ref{n-psi-redef-bc-d-alpha})
and
(\ref{n-A-decomposition-bc-d-alpha})
do not immediately allow us to write down the correlation functions
because, as in section~\ref{sec-n-vectors},
$M_0^2$ may be complicated, but
we can write\footnote{The
$I_f$ is unnecessary, but it is included to make the rest of the notation
below more transparent.}
\begin{equation}
M_0^2=M^2-\frac{1}{\pi}e\,e^T
=M\left(I-\frac{1}{\pi}M^{-1}eI_fe^TM^{-1}\right)M
\end{equation}
where $I_f$ is the identity matrix in the $n_F$ dimensional space of the fermions.
We should be able to find an inverse of the form
\begin{equation}
M_0^{-2}
=M^{-1}\left(I+\frac{1}{\pi}M^{-1}e
G_f
e^TM^{-1}\right)M^{-1}
\label{m0inversegf}
\end{equation}
Where $G_f$ is an $n_F\times n_F$ matrix.
What we need is (and this should remind you of (\ref{gamma-condition}))
\begin{equation}
\begin{array}{c}
\displaystyle
M_0^2M_0^{-2}=I
=M\left(I+\frac{1}{\pi}
M^{-1}e
\Bigl(G_f-I_f-\frac{1}{\pi}e^TM^{-2}e\,G_f\Bigr)
e^TM^{-1}
\right)M^{-1}
\\
\displaystyle
=M\left(I+\frac{1}{\pi}
M^{-1}e
\Bigl(G_f-I_f-B_fG_f\Bigr)
e^TM^{-1}
\right)M^{-1}
\end{array}
\end{equation}
which is satisfied if
\begin{equation}
G_f=(I_f-B_f)^{-1}
\label{gf}
\end{equation}
where the symmetric $n_F\times n_F$ matrix $B_f$ is defined by
\begin{equation}
 B_f \equiv 
\frac{1}{\pi}
e^TM^{-2}e
\label{bf}
\end{equation}
The matrix $M_0^2$ will be positive as long as the eigenvalues of the matrix 
\begin{equation}
\tilde B_f\equiv \frac{1}{\pi}
M^{-1}e\,e^TM^{-1}
\end{equation}
are less than one.  When the largest eigenvalue goes to one, we go to the
analog of the Schwinger point.  

\begin{equation}
\frac{1}{\pi}
M^{-1}e\,e^TM^{-1}\,v=v
\end{equation}
Acting on the left of both sides with $e^TM^{-1}$ gives
\begin{equation}
B_f\,e^TM^{-1}\,v=e^TM^{-1}\,v
\end{equation}

This correspondence holds for all non-zero eigenvalues.  
If there is a non-zero $v$ for which
\begin{equation}
\frac{1}{\pi}
M^{-1}e\,e^TM^{-1}\,u=\lambda\,u
\label{meeme}
\end{equation}
with $\lambda\neq0$, then $e^TM^{-1}\,u$ cannot vanish (because that would
imply $\lambda=0$) so that
acting on the left of both sides with $e^TM^{-1}$ gives
\begin{equation}
B_f\,e^TM^{-1}\,u=\lambda\,e^TM^{-1}\,u
\label{bfe}
\end{equation}
Thus very non-zero eigenvalue of $M^{-1}e\,e^TM^{-1}$ is an eigenvalue of
$B_f$.  A similar argument shows that
every non-zero eigenvalue of $B_f$ is an eigenvalue of
$M^{-1}e\,e^TM^{-1}$.

When we use (\ref{n-psi-redef-bc-d-alpha}),
(\ref{n-A-decomposition-bc-d-alpha}), (\ref{m0inversegf}), and (\ref{gf})
to write down the correlation functions, some useful simplifications
occur.  As we note in section~\ref{sec-n-vectors},  the $\cV$ and $\cC$
propagators are proportional to the identity in the gauge boson space.  The
implies, for example, that $M_0^{-1}$ appears in the fermion correlators
only in the combination
\begin{equation}
\begin{array}{c}
\displaystyle
e^TM_0^{-2}e
=e^TM^{-2}e+\frac{1}{\pi}e^TM^{-2}eG_fe^TM^{-2}e
\\
=\pi(B_f+B_f(I-B_f)^{-1}B_f)=\pi\,B_f(I-B_f)^{-1}
\end{array}
\label{bf-i-bf}
\end{equation}
In correlators involving a single $A^\mu_j$, a similar simplifcation
obtains because the $e^T\cdot M_0^{-1}$ in (\ref{n-psi-redef-bc-d-alpha})
is multiplied by and $M_0^{-1}$ from (\ref{n-A-decomposition-bc-d-alpha})
and correlator involves only the combination
\begin{equation}
\begin{array}{c}
\displaystyle
e^TM_0^{-2}
=e^TM^{-2}e+\frac{1}{\pi}e^TM^{-2}eG_fe^TM^{-2}
\\
=\pi(B_f+B_f(I-B_f)^{-1}B_f)=\pi\,(I-B_f)^{-1}e^T
\end{array}
\label{i-bfet}
\end{equation}

\section{Correlation functions\label{sec-g1}}

In the notation of section~\ref{sec-na-vectors-nf-fermions}, the fermion
correlation functions in generalized Sommerfield model with $n_A$ vector
bosons and $n_F$ fermions to infinite order in perturbation theory
are written explicitly below.
The most general fermion correlation function involves $n_{1\alpha}$
left-handed fermion and anti-fermions of type $\alpha$, 
$\psi_{\alpha1}(x_{1\alpha j})$ 
and $\psi^*_{\alpha1}(y_{1\alpha j})$ 
where $j=1$ to $n_{1\alpha}$
and $n_{2\alpha}$
right-handed fermions and anti-fermions of type $\alpha$, 
$\psi_{\alpha2}(x_{2\alpha j})$ 
and $\psi^*_{\alpha2}(y_{2\alpha j})$ 
where $j=1$ to $n_{2\alpha}$.
We find
\begin{equation}
\Braket{0|T\,
\left(
\prod_{\alpha=1}^{n_F}
\prod_{j=1}^{n_{1\alpha}}\psi_{\alpha1}(x_{1\alpha j})
\,\psi_{\alpha1}^*(y_{1\alpha j})
\right)\,
\left(
\prod_{\alpha=1}^{n_F}
\prod_{k=1}^{n_{2\alpha}}\psi_{\alpha2}(x_{2\alpha k})
\,\psi_{\alpha2}^*(y_{2\alpha k})
\right)
|0}
\label{general}
\end{equation}
\begin{equation}
= \left(\prod_{\alpha=1}^{n_F} i^{n_{1\alpha}+n_{2\alpha}}
 \,(-1)^{\frac{n_{1\alpha}(n_{1\alpha}-1)+n_{2\alpha}(n_{2\alpha}-1)}{2}}\right)
\label{csigns}
\end{equation}
\begin{equation}
\times\left(\prod_{\alpha =1}^{n_F}\prod_{\beta =1}^{n_F}
\prod_{j=1}^{n_{1\alpha}}
\prod_{k=1}^{n_{1\beta}}
C^{\alpha\beta}_0(x_{1\alpha j}-y_{1\beta k})\,C^{\alpha\beta}(x_{1\alpha j}-y_{1\beta k})\,S^{\alpha\beta}_1(x_{1\alpha j}-y_{1\beta k})\right)
\label{cfirst}
\end{equation}
\begin{equation}
\times
\left(\prod_{\alpha=1}^{n_F}\prod_{\beta=1}^{n_F}
\prod_{j=1}^{n_{2\alpha}}
\prod_{k=1}^{n_{2\beta}}
C^{\alpha\beta}_0(x_{2\alpha j}-y_{2\beta k})
\,C^{\alpha\beta}(x_{2\alpha j}-y_{2\beta k})\,S^{\alpha\beta}_2(x_{2\alpha j}-y_{2\beta k})\right)
\end{equation}
\begin{equation}
\times
\left(
\prod_{\alpha\leq\beta,\;j<k\; {\rm for}\;\alpha=\beta}
C^{\alpha\beta}_0(x_{1\alpha j}-x_{1\beta k})^{-1}\,
C^{\alpha\beta}(x_{1\alpha j}-x_{1\beta
k})^{-1}\,S^{\alpha\beta}_1(x_{1\alpha j}-x_{1\beta k})^{-1}
\right)
\end{equation}
\begin{equation}
\times
\left(\prod_{\alpha\leq\beta,\;j<k\; {\rm for}\;\alpha=\beta}
C^{\alpha\beta}_0(x_{2\alpha j}-x_{2\beta
k})^{-1}\,C^{\alpha\beta}(x_{2\alpha j}-x_{2\beta
k})^{-1}\,S^{\alpha\beta}_2(x_{2\alpha j}-x_{2\beta k})^{-1}
\right)
\end{equation}
\begin{equation}
\times
\left(\prod_{\alpha\leq\beta,\;j<k\; {\rm for}\;\alpha=\beta}
C^{\alpha\beta}_0(y_{1\alpha j}-y_{1\beta k})^{-1}\,
C^{\alpha\beta}(y_{1\alpha j}-y_{1\beta
k})^{-1}\,S^{\alpha\beta}_1(y_{1\alpha j}-y_{1\beta k})^{-1}
\right)
\end{equation}
\begin{equation}
\times
\left(\prod_{\alpha\leq\beta,\;j<k\; {\rm for}\;\alpha=\beta}
C^{\alpha\beta}_0(y_{2\alpha j}-y_{2\beta
k})^{-1}\,C^{\alpha\beta}(y_{2\alpha j}-y_{2\beta
k})^{-1}\,S^{\alpha\beta}_2(y_{2\alpha j}-y_{2\beta k})^{-1}
\right)
\end{equation}
\begin{equation}
\times
\left(\prod_{\alpha=1}^{n_F}\prod_{\beta=1}^{n_F}
\prod_{j=1}^{n_{1\alpha}}
\prod_{k=1}^{n_{2\beta}}
C^{\alpha\beta}_0(x_{1\alpha j}-y_{2\beta k})\,C^{\alpha\beta}(x_{1\alpha j}-y_{2\beta k})^{-1}\right)
\end{equation}
\begin{equation}
\times
\left(\prod_{\alpha=1}^{n_F}\prod_{\beta=1}^{n_F}
\prod_{j=1}^{n_{2\alpha}}
\prod_{k=1}^{n_{1\beta}}
C^{\alpha\beta}_0(x_{2\alpha j}-y_{1\beta k})\,C^{\alpha\beta}(x_{2\alpha j}-y_{1\beta k})^{-1}\right)
\end{equation}
\begin{equation}
\times
\left(\prod_{\alpha=1}^{n_F}\prod_{\beta=1}^{n_F}
\prod_{j=1}^{n_{1\alpha}}
\prod_{k=1}^{n_{2\beta}}
C^{\alpha\beta}_0(x_{1\alpha j}-x_{2\beta k})^{-1}\,C^{\alpha\beta}(x_{1\alpha j}-x_{2\beta k})\right)
\end{equation}
\begin{equation}
\times
\left(\prod_{\alpha=1}^{n_F}\prod_{\beta=1}^{n_F}
\prod_{j=1}^{n_{1\alpha}}
\prod_{k=1}^{n_{2\beta}}
C^{\alpha\beta}_0(y_{1\alpha j}-y_{2\beta k})^{-1}\,C^{\alpha\beta}(y_{1\alpha j}-y_{\beta y2k})\right)
\label{clast}
\end{equation}
where for the Sommerfield model
\be
C^{\alpha\beta}(x) =\exp\left[\sum_{j=1}^{n_A}\frac{e_{j\alpha}e_{j\beta}}{2\pi m_j^2}
\left[K_0\left(m_j\sqrt{-x^2 + i\epsilon}\right) + \ln\left(\xi m_j\sqrt{-x^2 +
i\epsilon}\right)\right]\right]
\label{C(x)}
\ee
\be
C^{\alpha\beta}_0(x) = \exp\Bigl[i\pi[B_f(I-B_f)^{-1}]_{\alpha\beta}\left[D(x) -
D(0)\right]\Bigr]
\propto \left(-x^2+i\epsilon\right)^{-[B_f(I-B_f)^{-1}]_{\alpha\beta}/4} 
\ee
with 
\begin{equation}
\xi= \frac{e^{\gamma_E}}{2}
\label{xi}
\end{equation}
 and
where the matrix $B_f$ from (\ref{bf}) can be written as
\begin{equation}
[B_f]_{\alpha\beta}=\sum_j\frac{e_{j\alpha}e_{j\beta}}{\pi m_j^2}
\label{bf-alpha-beta}
\end{equation}
{\renewcommand{\arraystretch}{1.4}\begin{equation}
S^{\alpha\beta}_1(x)= \left\{
\begin{array}{l}
S_1(x)\mbox{~~if $\alpha=\beta$}\\
1\mbox{~~if $\alpha\neq\beta$}
\end{array}\right.
\end{equation}}
\begin{equation}
S_1(x)
=\int \frac{d^2p}{(2\pi)^2}\,e^{-ipx}\,
\frac{p^0+ p^1}{p^2+i\epsilon}
=-\frac{1}{2\pi}\frac{x^0+
x^1}{x^2-i\epsilon}
\label{s1}
\end{equation}
{\renewcommand{\arraystretch}{1.4}\begin{equation}
S^{\alpha\beta}_2(x)= \left\{
\begin{array}{l}
S_2(x)\mbox{~~if $\alpha=\beta$}\\
1\mbox{~~if $\alpha\neq\beta$}
\end{array}\right.
\end{equation}}
\begin{equation}
S_2(x)
=\int \frac{d^2p}{(2\pi)^2}\,e^{-ipx}\,
\frac{p^0- p^1}{p^2+i\epsilon}
=-\frac{1}{2\pi}\frac{x^0-
x^1}{x^2-i\epsilon}
\label{s2}
\end{equation}

The ten contributions in (\ref{cfirst})-(\ref{clast}) correspond to the ten
lines in the diagram
in figure~\ref{fig-1}. (\ref{csigns}) includes a factor of \(i\) for every
free-fermion correlator and factors of \((-1)^{n(n-1)/2}\) from combining 
the different contractions of left-handed and right-handed free-fermions into a single factor. 
{\figsize\begin{figure}[htb]
$$\includegraphics[width=.7\hsize]{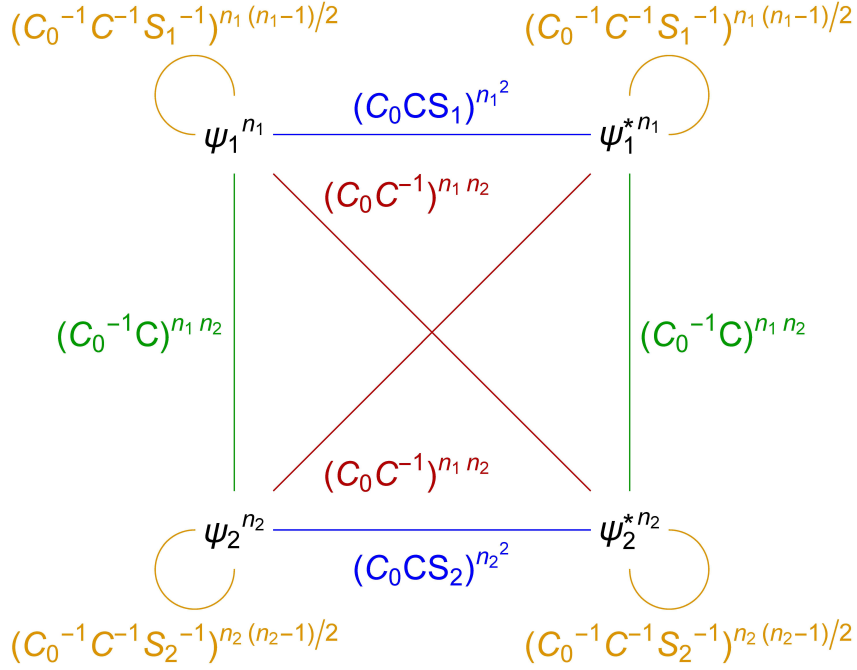}$$
\caption{\figsize\sf\label{fig-1}Pictorial representation of the fermion
correlation functions}\end{figure}}.  

\section{Conclusions\label{sec-conclusions}}

The generalizations of the Sommerfield model analyzed here are a rich
vein of exactly solvable quantum field theories some of which can exhibit
properties that are very different from the those of the simple Sommerfield
and Schwinger models.  We look forward to mining this lode in future work.

\section*{Acknowledgements\label{sec-ack}}

This work is supported in part by NSF grant
PHY-1719924.  BW's research is supported in part by the Harvard College
Research Program.

\bibliography{up4}

\end{document}